# Roadmap on Material-Function Mapping for Photonic-Electronic Hybrid Neural Networks

Mario Miscuglio[1,†], Gina C. Adam[1,†], Duygu Kuzum[2], Volker J. Sorger[1,*]

[1]George Washington University, 800 22nd Street NW, Washington, DC 20052, USA
[2]University of California, San Diego, Lo Jolla, CA 92093, US
†Both authors contributed equally to this manuscript.
*Email: sorger@gwu.edu

**Abstract—** The state-of-the-art hardware in artificial neural networks is still affected by the same capacitive challenges known from electronic integrated circuits. Unlike others emerging electronic technologies, photonics provides low-delay interconnectivity suitable for node-distributed non-Von Neumann architectures relying on dense node-to-node communication. Here we provide a roadmap to pave the way for emerging hybridized photonic-electronic neural networks by taking a detailed look into a single node's perceptron. We discuss how it can be realized in hybrid photonic-electronic heterogeneous technologies. Further, we assess that electro-optic devices based on phase change or strong carrier dispersive effects could provide a viable path for both the perceptron 'weights' and the nonlinear activation function in trained neural networks, while simultaneously being foundry process-near materials. The study also assesses the advantages of using nonlinear optical materials as efficient and instantaneous efficient activation functions. We finally identify several challenges that, if solved, could accelerate the adoption of such heterogeneous integration strategies of emerging memory materials into integrated photonics platforms for near real-time responsive neural networks.

## 1 Introduction:

While the scientific community still lacks a full understanding of the operation of the human brain, we yet can draw some parallels to compute-systems with respect to operating efficiency. Machine learning (ML) tasks performed by neural networks (NN) can be used for such computer-vs.-brain comparison, since both are examples of nonlinear hypothesis systems that can be trained to classify patterns. Interestingly, one of the fastest computers[1] is only able to simulate ~1% of human brain activity in requiring about one hour and demanding massive compute infrastructure overheads (e.g. 82,000 processors, and 1.73 billion virtual nerve cells connected by 10.4 trillion synapses) consuming about 10 megawatts, while the human brain operates on 10's of Watts[2].

In the aim to compensate this disparity and to match some of the brain's computational and energy efficiency, the recent efforts to develop non-Von Neumann neuromorphic hardware are capable of efficiently implementing artificial NN operations, in particular, vector matrix multiplication (VMM) and backpropagation[3–5].

Conventional (Von Neumann) compute architectures utilizes centralized processors, and relies on logic technologies, while execute stored programs sequentially. On the other hand, emerging non-Von Neumann systems are inherently i) decentralized and therefore ii) require significant communication between these many nodes of the network, yet iii) process information in parallel and iv) are trainable for applications such as in machine learning.

Electronic implementations of such brain-mimicked systems show significant (3-4 orders of magnitude[6,7] energy-per-compute reductions defined by the underlying mathematical interconnection function of NNs, namely, multiply-and-accumulate (MAC). Thus, the efficiency and performance in the number of processed MAC's per unit time and unit energy (i.e. MAC/s and MAC/J) are two key metrics to be considered when new hardware technology and emerging materials are explored for NN technology.

While electronic NN-based processors have shown improvement with respect to energy efficiency, the time delay to obtain a classification result from the NN (e.g. of an inference ML task) is not improved as compared to regular Von Neumann systems, due to the high-RC delay of electronic circuits.

In this work we consider the NN being trained offline, and hence only discuss the performance under inference operation. This approach, yet, requires mapping the ML-task to the NN hardware, which is not straightforward for photonics networks at the time this paper is written. However, initial approaches of modifying graph-simulators (e.g. Tensorflow) to include physical effects from the device layer (e.g. impact of noise, cascadability) show higher inference accuracy using electro-optic hybrid approaches for NNs.

Emerging memory technologies, such as phase-change memory[8] (PCM), conductive bridge random access memory[9] (CBRAM), resistive switching memory[10] (RRAM), have been demonstrated to be compelling candidates as synaptic devices for weight storage and matrix vector multiplication (VMM) in purely electronic neuromorphic circuitry. These devices are characterized by exceptional characteristics such as high footprint scalability, multi-bit storage capability, and long retention-time non-volatility as well as an overall higher technological maturity compared to integrated photonics. Therefore, few efforts[11–15] have targeted integrating non-volatile materials with waveguides, in order to explore novel ways of tuning their refractive index, either plasmonically of the optical mode or through phase changes in the crystallinity of the material. From this perspective, here we argue that integrated photonics and memristive hybrid systems (**Fig. 1**) could present significant improvements to the established digital implementations of VMM-tasks using graphic- and tensor process units (GPU and TPU). However, their usage aiming to replace the current technology is still hindered by the abovementioned hurdles and the state of the community of integrating CMOS compatible phase change materials[16], such as GST ($Ge_2Sb_2Te_5$), with PIC is relatively at its infancy.

Nevertheless, the integration of the two technologies seems to be an interesting direction to pursuit to complement their strengths (**Fig. 1a**). Yet, if successful, such hybrid integration would be particularly appealing since it could enable retention of the optical information, digitalization of the output, and introduce nonlinearities at the same time, while building on existing process know-how and made capital investments (i.e. material-foundry compatibility). The underlying mechanism for achieving weighting functionality in integrated photonics using memristors relies on specific active materials[12,14,16–20] either alone (weighting + storing) or in combination with electro-optic modulators (storing), aiming to alter their optical response in a non-volatile reversible way, by means of an activation potential (i.e. thermal, electrical or optical).

With the aforementioned electronic-NN based ML successes, the question becomes why perform NN's in the optical domain? Before we answer this, one can more generally ask what the rationale might be for performing information processing in optics and in integrated photonics? Here we argue that the main reason for photonic NN hardware is found in the short delay[21], which could be important for applications such as in ranging, synthetic aperture radar, automated target recognition, or nonlinear predictive control[22] (**Table 1**). Once the dot-product, providing the 'weights' of the photonic perceptron, are set (i.e. trained offline), the entire system only depends on the time-of-flight of the photon through the NN plus the response time of the electro-optic components of the perceptron[5,23–25]; here several implementation options exist, and for the discussion of this roadmap we focus on the dot-product photonic weights and the nonlinear (NL) activation function. That is, the entire NN's delay (once trained) can be 100's picosecond short using integrated photonics and high-speed (10's GHz) photodetectors. In addition to delay, there are other benefits using photonics for NNs or well-established information processing paradigms such as parallelism from WDM and vanishing capacitive 'wire' delay (**Table 1**). However, challenges exist too such as in function-per-wafer-footprint, electro-optic (EO) conversion efficiencies, packaging, and – most relevant for this this roadmap paper – efficiency and number of states of photonic memories, whom development is still in its beginning. It is this very reason that we turn our attention to a hybridization strategy by combining the best-of-both-worlds (**Fig 1a**); that is, using the non-volatility from electronics yet the interconnectivity from optics combined with the compactness from integrated photonics which also aims to leverage semiconductor process economy of scale, through foundry services such as AIM, IMEC or IME.

While a number of both fundamental and practical strengths exist, there are realistic 'pain points' as well, that need to be overcome in the long term in order for photonics (i.e. integrated optics) to become competitive. Though, we note also that details on hardware competitiveness depend strongly on the applications sought after.

While volatile memory options in photonics could be implemented for the weighting function using, for example, electro-optic modulators[26–33] their high optical material index drifts, due to thermal and electrical noise, is sub-optimal for the typical longevity of NN weights while performing inference of unseen data, which change only when a new training set is obtained (if ever), which can be particularly time consuming due to the enormous amount of training data employed to ensure correct training in a high dimensionality problem, namely the 'curse-of-dimensionality'[34]. Therefore, we firmly believe that the community should explore emerging materials that feature favorable performances for photonic, electro-optic material performance while keeping chip integration synergies in mind, as discussed below.

| Rationale for Photonic Information Processing | |
| --- | --- |
| **STRENGTHS** | **PAIN POINTS** |
| • Nanosecond Delay → 'Real Time' | • Efficient Memory Functionality |
| • Bosonification → Parallelism (e.g. WDM) | (Low IL, Efficiency, Multistate) |
| • Low loss communication → 'Wire' is free | • Footprint |
|   →Distibuted Non von-Neumann Architectures | • E-O Conversion |
| • 'One-shot' Execution (Non-iterative) | • Packaging/Alignment |
| • Fabrication in progress → Aim Photonics, IMEC, IME | |

*Table 1. Rationale to perform information processing such as exploited in artificial neural networks for machine-learning tasks in the optical domain.*

Neuromorphic photonics and memristor-based systems provide solutions that can deliver particularly high performances in terms of either throughput or energy efficiency by comparison with the current purely CMOS based execution[25,35]. The main advantage of exploiting Photonic Integrated Circuit (PIC) for NN tasks is intrinsic to the wave-nature of the signal that carries information (**Fig. 1a**). The fundamental operation in NN-tasks are multiplications and accumulations (MAC), which can be performed without any additional energy because a MAC is achieved by simple interference of phase-shifted electromagnetic waves travelling within waveguides[5,24,25]. Moreover, integrated photonics is naturally predisposed to ultra-high speed characterized by short latency due to a) the time-of-flight of the photon in the chip (few ps for large chips) and b) to a considerable extent by its electro-optic components (few tens of ps), such as detectors and modulators. Furthermore, exploiting Wavelength Division Multiplexing (WDM)[5], PICs can perform selective weighting operation on multiple inputs mapped on different wavelengths, using the same physical channel, supporting higher parallelism and remarkable throughput[5,36]. Furthermore engineering light-matter interactions enables advances in efficient modulators, which can realize 100's of attojoule-per-operation energy dissipation[37–39].

Nevertheless, the advancement of integrated photonics in neural networks is hindered by major challenges; beside the larger footprint with respect to microelectronics, which could be neglected considering the higher throughput and efficiency, the advancement of PICs at a larger scale is primarily held back by difficulties in the integration with a CMOS interface, local laser sources and packaging (**Table 1**). This is because the materials employed require dedicated industrial process recipes that have not reached enough maturity yet to provide cost effective, consistent, or scalable results. Just to clarify this point further: PICs-based components and sub-systems are commercially sold in the millions namely for data-centers, but using PICs for NN is a technology direction which is still at its infancy despite great progress having been made [5,7,22,23,25]. It also worth mentioning that, except for recent efforts[23], NNs in integrated photonics are still lacking a straightforward implementation of an optical nonlinear activation function, to mimic the action potential firing in the neuron or an analog tuning such as favored by the computer science community (e.g. rectifying linear unit ReLU). Also, the absence of a straightforward non-volatile memory realized in photonics that can be written, erased, and read optically still limits the realization of all-photonic chip-scale information processing for NN tasks, yet initial designs concepts exist[24]. What we find is that the required power consumption and bandwidth could both be improved when non-volatile memory functions could be integrated directly in the optic domain, which is particularly advantageous in trained networks when the weights are fixed (or changing seldom).

Our aim in this work is to explore a roadmap of non-volatile materials and approaches that can be integrated with photonics to enable dot-product multiplications and thresholding in the optical domain. Here, we review, discuss, and project challenges and opportunities for device type and material choices for these memristor-based electronic-photonic hybrid NNs from the perspective of key metrics such as power dissipation, electrical vs. optical readout, multi-state tunability to mention a few, thus laying a corner stone towards establishing a materials roadmap for photonic NNs. We discuss the advantages and issues related to a hybrid integration of photonics and memristors in a complementary way, highlighting material systems, which are jointly shared between the two technologies and that could allow a seamless integration. CMOS compatibility and large-scale integration pose additional material-related constraints, limiting the choice of materials available for electrodes, device layers and isolation.

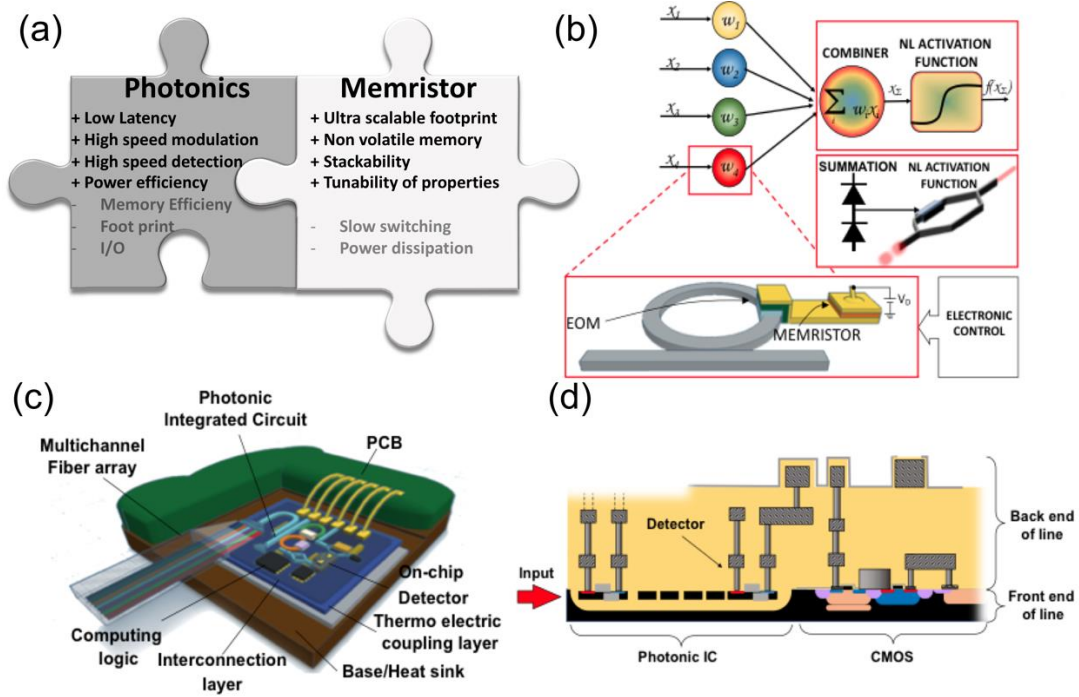

*Figure 1. Hybrid memristive photonic systems towards building dual-technology neural network (NN) architectures. (a) Specific pros and cons which characterize the two technologies, integrated photonics (dark) and memristor (light). (b) The perceptron model representing a single node of a NN can be mapped onto photonic hardware components. Thus once the NN is trained and the weights are set in the electro-optic weighting modulators, the delay inference for the inference task is given by the photons time-of-flight through the photonic integrated circuit (PIC), which can be taken to be 'real-time' (i.e. 1-10's ps) compared to electronic NN solutions. However, the opto-electronic weights require a constant voltage bias taking a toll on the power consumption. Thus, real-time fast and power efficient photonic NNs can be designed by integrating non-volatile memory elements near the photonic weight components. This is synergistic to offline trained 'weights', which are updated infrequently, if ever. (c) Schematic of a back and front end-of-line of hybrid integration for non-volatile weights. In addition to technological performance gains, the material compatibilities with fabrication processes of foundries need to be considered as well. (d) Schematic of a purely illustrative monolithic Silicon or $Si_3N_4$ photonic integration, which comprises planar integration of Si photonics with complementary metal oxide semiconductor (CMOS) circuits on the same layer.*

## 2   INTEGRATED PHOTONICS FOR NEUROMORPHICS

For hybrid memristor-photonics NN systems, the fast and efficient response of the transfer function of an electro-optic modulator is mostly suitable for mimicking the NL activation function of the neuron rather than the weighting, since the network is mostly offline trained and performing only the inference tasks[5,23,40]. On the contrary the ability of the memory element to retain information for a comparable long time, can be exploited for the purpose of efficiently storing the weights and modulating the signal travelling in the waveguide accordingly. In this section (Sec. 2.2), we discuss recent work on integrated photonics platform, optical modulators and their integration with CMOS electronics, before we further discuss key metrics of the state-of-the-art modulators.

## 2.1 Passive Photonic Neural Network Interconnectivity: waveguide platform options

For PIC operating at telecommunication wavelengths (1550 nm) the main platform for planar light-wave circuit is Silicon on Insulator (SOI). The crystalline silicon layer atop the insulator is used to create optical waveguides (through optical index contrasts) and can be extended to include both passive and active devices used to deliver NN functionality. The SOI chip can be realized by either smart cut or SIMOX processes. The buried insulator enables propagation and strong confinement of infrared light in the silicon layer on the basis of total internal reflection, with low propagation losses (<1 dB/cm)[41] and small bending radii (<5μm)[42] enabling PIC with compact footprint. This SOI platform also features monolithic electro-optic modulators, i.e. without the addition of any other material, either by thermally or electrostatically changing Silicon's optical refractive index[43–46]. Silicon's optical (i.e. all-optical) nonlinearities arise at few tens of mW, which is a value that could limit the depth (numbers of layers) of the NN, and contemporary sets the limits to the insertion losses admissible.

The integration with CMOS electronics for logic circuitry with SOI, while technologically feasible[47], is challenging due to technical and economic mismatches, therefore hybrid integration is usually required. Layer-stacking and integration with light sources could be feasible by SOI wafer bonding techniques such as coupling the evanescent optical mode of a III-V laser to SOI waveguide[48,49], high-performance quantum dot (QD) lasers monolithically grown on Si[50,51], or other electrically driven solution[52]. While Silicon's bandgap is transparent at telecommunication frequencies, there is no conceptual rule why PIC-based NN could not operate at visible or also at mid-IR wavelengths. In fact, there are convincing reasons to consider small wavelengths; a) the smaller wavelength enables denser PICs, b) the higher bandgap can deliver lower optical losses saving chip power consumption, and c) extending the pump-power range before NL become parasitic[20], thus enabling NN cascadability. Such a visible-photonics platform material is $Si_3N_4$, which presents a bandgap at higher energy (0.4μm). Interestingly, the process recipes for silicon nitride are favorable for large-scale photonic networks; when deposited by LPCVD $Si_3N_4$ is characterized by both high material stability and refractive index regularity, as well as nanoscale etch-resolution and lower surface roughness, which leads to reduced scattering and hence low optical losses per unit length (<10 dB/m) for a 0.5 mm bending radii, which is 10-100x lower compared to SOI.[53] In contrast to SOI, due to the fundamental flexibility of the deposition techniques[54], $Si_3N_4$ also allows for easy integration and 3D stacking flexibility with either SOI or CMOS platforms[55]. Regarding other material integration into this silicon nitride platform, several options have been successfully embedded such as a number of metals or colloidal quantum dots[56]. This platform also allows creating more complex photonic structures such as distributed feedback reflectors (DFB)[57], and has shown to deliver PICs with multiple photonic layers. Regarding electro-optic (EO) phase tunability, silicon nitride is less responsive than even silicon and hence is not a suitable material for monolithic weights of photonic NNs. Nonetheless, given its promising passive properties, it appears an ideal material for heterogeneous integration in addition to co-integration with SOI and CMOS logic circuitry.

## 2.2 Electro-optic 'Weights' and Nonlinear Activation Function: materials for high-speed and efficient modulators

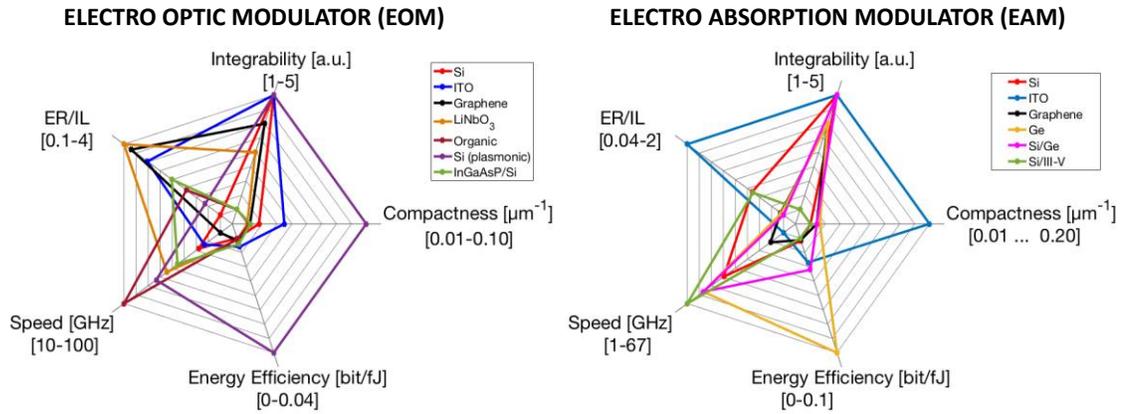

*Figure 2.* Electro-optic modulators can be used as efficient tunable 'weights' of the perceptron, but also as the nonlinear (NL) activation function 'threshold'. For the latter the 'weighted-addition' signal can be provided by a photodetector, whose photo-generated voltage becomes the input 'gate' bias of the modulator. Refer to Ref.[7,22,23] for details on possible perceptron architectures or to Fig. 1b. Comparison of key metrics for electro-optic and absorption modulators for different material systems. The figure of merit considered are the Integrability (qualitative number 1 low to 5 high), Compactness (1/Length expressed in µm), Energy Efficiency (bit/fJ), Speed (GHz) and modulation dynamics vs losses (ER/IL refers to the ratio between extinction ratio and insertion loss), The values of the key metrics increase moving further away from the center of the radar chart. Silicon[58], ITO , Graphene[27], LiNbO3[26], Hybrid-Organic[33], Si-plasmonic[29], InGaAsP/Si[60] Electro-optic modulators are reported (left). Silicon , ITO[32], Graphene[62], Germanium[63], Si/Geraminum[64] , III-V/Si[61] Electro absorption modulators (right). Integrability is an arbitrary value (1-5), given based on the "standardization" of the fabrication process as well as compatibility of the material with CMOS technology." CMOS compatible processes such as Si and (now recently also) ITO based modulators received 5 points, foundry-close materials (Ge, Si/Ge) and processes 4 points, materials that show R&D-level evidences defined as successful material co-integration into pre-taped-out chips, such as integration with SOI received 3 points (Graphene, LiNbO$_3$), 2 points are given to materials that are still relegated to the research level and 1 star to materials that are not compatible with standard CMOS processes (InP and polymers) and require complex integration or limited by small scale integration capability."

To modulate the light-wave information travelling in the waveguides, efficient electro-optic or absorptive modulators must be designed and integrated, either monolithically or heterogeneously in the SOI or Si$_3$N$_4$ platform. Modulators are active photonic components that induce an optical absorption (EA, electro-absorption), or change the optical path length or refractive index (EO, electro-refraction) of a material[39]. The choice of Electro-optic modulation over electro-absorption and vice versa is merely related to the device type and its use in the network. Using either electro-refractive or electro-absorptive modulation ultimately translates into an intensity modulation. Phase modulators require interferometric scheme to induce an amplitude change, such as the Mach-Zehnder (MZ) configuration or Micro-ring Modulators (MRM), whereas absorption modulators achieve it in a linear waveguide geometry. MRMs and MZM operate at specific wavelength, while electro-absorption modulation are principally broadband. MRMs can be employed in architectures that exploits the available bandwidth using Wavelength Division Multiplexing (WDM), in which WDM inputs are weighed through tunable MRRs multiplexed on the same physical channel. In the end, a modulator is a multi-trade-off device, and a straightforward comparison without application context lacks relevance. For instance, the material system LiNbO$_3$ has shown lower power and high-speed performance[26,65], but comes with very large footprints (~cm size) due to the lack of any resonance effect. The other extreme is, for example, a Graphene-plasmon modulator[37,66] deploying both optical and material resonance allowing for micrometer small devices, yet at a (much) higher per-unit length device insertion loss. In fact, we have explored such material-mode-performace options for a variety of EO-active materials (Si, ITO, Graphene, III-V, QD, TMDs) and multiple optical modes (photonic, plasmonic, hybrids) in ref [28,67].
Nevertheless, the performance of the electro-refractive/absorptive modulators depends on the physical modulation mechanism such as Pockels effect (LiNbO3), Kerr effect[68,69], quantum confined stark effect[70] (Ge-Si/Ge) and free carrier modulation ($Si^{44}$ and $ITO^{28}$).

For neuromorphic applications the optical modulator is a key component having a significant impact on the overall metric of the NN. Indeed EA/EO modulation can perform two of the three functionalities required in NN's rely on a high interconnectivity function between nodes, hence scaling questions rooted in component 'performance-per-overhead' should be considered. In fact, a limited modulation range can translate into a non-efficient type of NLAF, which would not allow to discriminate the data and ultimately hinder the accuracy of the network in the classification task, something we have shown in a recent paper[23], where we investigated the NN inference accuracy as a function of the actual EO modulator material, and where the modulator and photodetector. Taken together, both interconnectivity density and modulation performance are key for the core of the perceptron mechanism. The modulator can also be used for online learning by actively adjusting the weights of a to-be-trained $NN^{11,51}$. That is, when the network is trained offline, the weighting can be realized completely passive without energy consumption using EO or EA modulators that are gated via a non-volatile storage element or by photonic memories(**Fig. 1b**). Whereas, the NL activation function would mostly still rely on EO or EA modulators in combination with photodiodes. As reported by Miller[52], while providing a signal modulation depth (extinction ratio, ER, >3dB) at a sustained speed (>25 Gbit/s), the target power consumption for modulators has to be within few fJ/MAC (or better) for global on-chip connections to become technologically competitive (i.e. BER < $10^{-8}$). Materials such as silicon[43,72], lithium Niobate $(LiNbO_3)^{73}$, Germanium[30,63] and hybrid III-V[60,74] are regularly employed as active component in modulators and have showcased reliable, and to a certain extent, scalable results. Although, their use in densely integrated photonic circuit is not a viable approach due to their vastly (~$10^6$x) larger footprint compared with electronic switches (e.g. MOSFETs). Usually, for improving the extinction ratio while enabling relatively restrained footprint, these materials are embedded on resonant structures such as microring resonators, thus limiting the optical response. Considering the overall platform for PIC-based NNs, the most obvious material would be Silicon-based modulators for weights and/or NL activation, although as displayed in **Fig. 2**, Si-modulators based on carrier injection display low dynamic modulation (i.e. low ER) and if doped are particularly lossy (high insertion losses, IL), and additionally if thermally driven modulators are slow (1-10 kHz). On the other hand, even though characterized by competitive performances, III-V modulators are difficult to integrate into SOI platform as well as with CMOS circuitry, due to process and material incompatibility, and the wafer sources are costly (10's times compared to SOI). Instead, $LiNbO_3$ based modulators requires a 1-20 cm-large device lengths for inducing a π-shift if CMOS-compatible voltage biases (1-2V) are desired[26], thus hindering integration density.

In the view of providing modest energy reductions when performing NN tasks compared to electrical approaches, while still preserving a modest footprint, several engineering design and material choices must be made. As a main point, evidently, the selection of the active material is strongly impacted by its ability for voltage-efficient optical index tuning for the NL activation function and the weights; ideally, the absolute difference in the transmission of the modulator in its two states (ON/OFF), avoiding modulators with significant background loss, therefore not using those active materials which present high intrinsic losses in the OFF-state. Although, this might not be sufficient in order to achieve a sufficiently high modulation performance and low energy-per-compute for surpassing electronic efficiency, therefore an enhanced modulation can be reached by using either quantum-confined system or sub-diffraction limited plasmonic structures, either with monolithic or heterogeneous integration of other materials or structures. This results in a low-energy consumption of few attojoules-per-bit[37,39], up to the highest speeds[27,33,43,75], which corresponds to a compounded merit improvement of $10^5$ times compared to electronic switches. Hence, many research groups have strived for engineering on-chip modulators beyond the solutions offered by heterogeneous integrated photonic-foundries to date. Recent developments of monolithically and CMOS compatible integrated emerging EO or EA materials, such as Indium Tin Oxide (ITO)[28,32], graphene[27,31,76,77], quantum-confined structures[78] and TMDs into Si-photonics with specific device configuration aiming to enhance mode overlap allowed energy efficient[79,80], compact silicon photonic based modulators. The important performance metric for EOMs include high ER (>3dB), low IL (<1dB), modulation speed (>25 GHz), low energy consumption per bit (<10fJ/bit), and compact footprint area (possibly 3D volume). ITO can be particularly suitable for heterogeneous integration in Si exhibiting formidable electro-optic effect characterized by unity-order index change at telecommunication frequencies. ITO carrier-based electro-absorption models, which are implemented via capacitive gating[32], have shown sharp dynamic range, compact footprint and potential for GHz-fast modulation.

In recent works Amin et al[28] demonstrated a monolithically SOI-integrated ITO electro-optic modulator based on a Mach Zehnder interferometer (MZI) featuring a high-performance half-wave voltage and active device length product of $V\pi L = 0.52$V·mm. This device demonstrates a unity-strong index change in the active ITO layer enabling a 30 micrometer-short π-phase shifter, while purposefully operating ITO in the index-dominated region away from the epsilon-near-zero (ENZ) point, hence reducing optical losses. Moreover, some major electronic manufacturers have recently declared to integrate ITO into their foundry processes, hence it can now be termed to be a CMOS compatible material as it can be monolithically integrated in the photonic frameworks and directly interfaced with on-chip logic circuitry and memories (**Fig.2**).

ITO has the advantage over silicon of higher tunable absorption (in EAM scheme in proximity of ENZ[32,81] and unity variation of the refractive index (in EOM devices[28,82] away from ENZ). The inherent low tunability of Si under electrical bias causes inadequate performances with respect to ITO modulator given a fixed modulator length[83]. Moreover, by engineering ITO process one can tailor its optical response, thus increasing the modulation dynamic and lowering insertion losses, without necessitating plasmonic or dielectric cavities or ring resonators.

Incidentally, GST, discussed in the following section, is a second candidate that will enter foundry processes soon, while Graphene has yet to receive such 'permission', due to limitation in the substrates used for the growth and the inconvenience of the required transfer. In addition to ITO-based modulators[28,32,77], other active opto-electronic components are also demonstrated; Kim et al[59], for instance, experimentally demonstrated efficient and potentially high-speed directional coupler-based on ITO, that can also be employed as intensity modulation for NN weighting. Still at a research level, due to their non-scalable integration, other emerging materials such as Graphene[84,85], transition metal dichalcogenide flakes[85,86], organic[33] modulators and detectors have been reported, demonstrating, in some cases, striking performances yet process maturity is far-away from foundry standards. As aforementioned, integrated photonics necessitate materials which could provide memory functionality. Recent work on different device configurations has showcased capabilities of optically write/read/reset functionality as described in the next section.

Next, we proceed in the depicting the hybridization of photonic NN, discussing details of novel memory devices embedded in photonic framework, used for storage and weighting functionalities.

# 3 NOVEL MEMORY DEVICES FOR PHOTONIC INTEGRATION

In a hybrid photonic system, memristor devices would be used mainly as weights, which do not require high-frequency updating. Therefore, the efficiency of programming them during training (either directly on the photonic hardware or uploading offline-trained weights) is of importance in terms of chip power budget and compatibility of available voltage ranges (i.e. signal dynamic range). Recent studies[87–89] have highlighted the difficulties of training large-scale memristor networks acting as weights for neuromorphic computing. An alternative approach is to alleviate these challenges and to use both long- and short-term memory elements to implement these different temporal weight classes.

## 3.1 Efficient and Long-term Weights: heterogeneous integration of non-volatile memory

There are several long-term non-volatile memory technologies that show promise for integration with photonics systems **(Fig. 3)**. Given their technological maturity, flash memories (i.e. floating gate transistors) have the potential for gating the above-discussed modulators that perform the weighting or NL thresholding (**Fig. 3a**). However, in a typical configuration for regular memory applications, these transistors use polysilicon for the floating gate. For photonic integration, however, this option is not very desirable since the low-number of trapped charges can only impose slight changes in the modulators' refractive index (low dynamic range). Weight banks or NL activation functions based on such a technology option will have limited neuron bit-density limiting inference accuracy and NN cascadability. This essentially rules out online-learning options, where these modulators had to be adjusted according to a gradient-descent algorithm via back-propagation, which relies on differentiability of the modulators transfer function. For steeper dynamic range it is possible to achieve higher bit-density and consequently higher training efficiency, i.e. lower power. Also, the additional electrical capacitance from poly-depletion has increases both the device' RC-delay and the energy-per-MAC similar to arguments in transistor technologies and hence high-k dielectrics and metal-based floating gates may be a more viable path. Recent work[90] proposed a floating gate utilizing a non-volatile optical switch, where a graphene sheet is used as floating gate material[91] with much larger refractive index variation[31] (**Fig. 3a**). However, electron-based memory storage can be problematic from the perspective of increased optical loss in semiconductor waveguides, reduced non-volatility and limited analog programmability. A more desirable approach could be integrating memories heterogeneously with photonic waveguide-modulators. Such micrometer-compact device-to-device integration was recently pointed out by D. Miller[39] due to improved capacitive loading;

since every micrometer of metallic wire has a capacitance of 0.2fF, just a few micrometer of wire would bring the power budget above 1fJ-per-operation (i.e. bit or MAC) just for the wiring alone (excluding the functional devices) for a $V_{DD}$ of just 1 volt.

For such 'tight' integration, phase change memory (PCM) materials could be a viable option. Chalcogenide phase change materials based on germanium-antimony-tellurides (GST-PCMs) displayed remarkable properties in non-volatile memory technologies due to their high write and read speeds (ranging from few tens of ns to 1 ns[12]), high degree of integrability, CMOS compatibility, contained power consumption, long data retention and multi-level storage capability. In photonics, the GST-PCM can significantly change the effective refractive index of the waveguide by local amorphization or crystallization, which produces unity-strong index changes (i.e. $10^3$-$10^4$ times stronger than the free-carrier modulation of Silicon, for example).

Furthermore, PCMs have a state retention time of (estimated) >10 years and analog programmability up to 3-4 bits during the 'write' transition. Recent experimental demonstration[92] showed a photonic synaptic behavior of a chalcogenide GST phase-change film integrated with a $Si_3N_4$ waveguide (**Fig. 3b**). To achieve precise control of weight-programming using fixed-pulse characteristics, an innovative tapered waveguide structure with multiple discrete PCM islands was used, demonstrating 3bit (8 level) operation using ~400pJ for single pulse weighting. However, for such a power, if this GST-based synapse in a photonic multilayered perceptron framework are used for performing training they would not provide any significant advantage with respect to the state-of-the-art electronic neuromorphic systems or hybrid analog NVM-based approaches[87], therefore reducing the photonic advantages.

However, the short NN delay would still be the main rationale for photonic PCM-hybrid NN systems, when performing inference. To further clarify this concept, when the network is trained off-chip (i.e. in electronics e.g. GPU), the weights are set, and will (typically) not be updated often depending on the application (e.g. possibly weekly, monthly, yearly). On a device level, this means that the PCM-based memory must be written only once. This means that the optical phase-control of a portion of the PCM film must be altered only when the NN is set and should retain its phase throughout the inference process, in order to effortlessly modulate the quickly varying input signals, according to the value of the weights obtained by the off-chip training. In that instance, MAC operations are performed in a complete passive fashion, with the obvious energy benefits.

On the other hand, if the network is performing training on-chip, the limiting factor would be the update speed of the weights, which need to be constantly and ideally rapidly updated. Large and continuous amount of data fed to such a NN will be used for updating the weights through algorithms based on backpropagation and gradient descent. If the weight functionality is implemented with current photonic memories based on PCM the network would require a rather long learning time. Also, in terms of overall energy consumption when performing training, switching continuously the phase of the GST demands substantial amount of energy consumption (tens of pJ per weight update), which scales superlinearly with the number of nodes and layers. Therefore, during the training process, having fast and energy-efficient switching materials is not just desirable but compulsory. Current PCM switching energy (tens of pJ/bit) and delay (tens of ns) are at least 2 orders of magnitude larger than current electro-optic modulators and would not provide an advantage compared to electronic implementation when performing training on-chip. Replacement of GST with engineered phase-change materials that have low switching powers and fast response is therefore desirable from a materials roadmap perspective[93]

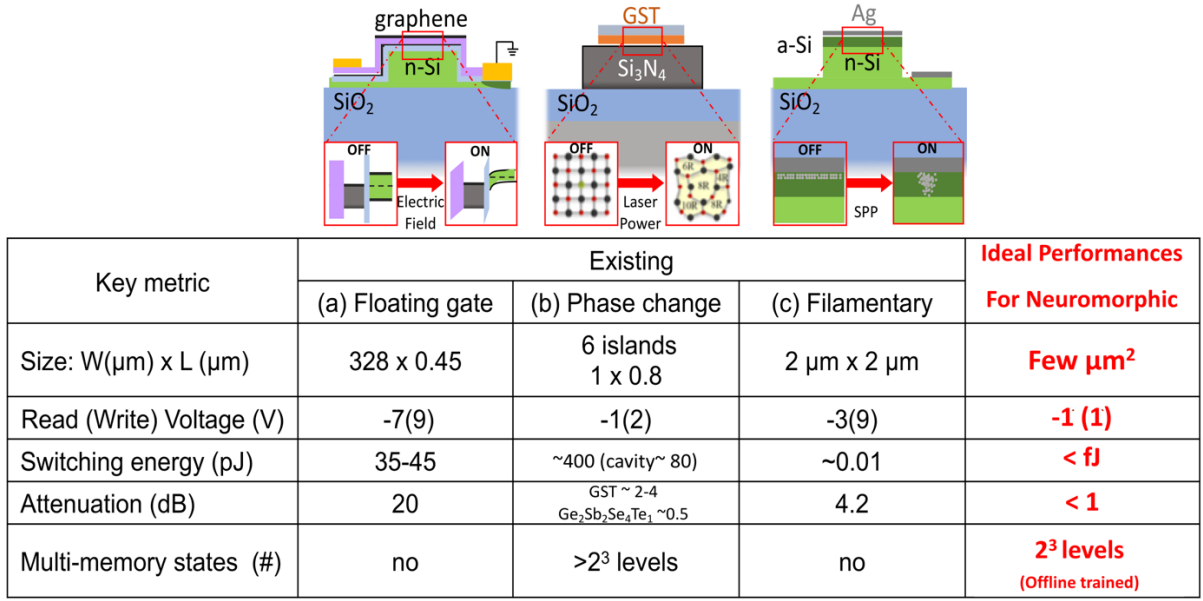

*Figure 3.* Designs and key metrics of memory technologies monolithically integrated with waveguides. (a) graphene-based floating gate[90]. (b) phase change memory synaptic device with optical readout [92] [Number of levels can exceed 3[94], Attenuation can be minimized using specific alloys [95], Switching energy reduced when integrated in cavities [96]]. (c) plasmonic memristor with Silver filaments [97]

Another intriguing alternative is plasmonic memristive devices that exploit the modulation of electron transmission in optical systems through filamentary switching (**Fig. 3c**). Memristive devices based on metallic or highly reduced oxide filaments can be used for this purpose, for example. Plasmonic memristors based on silver filaments and integrated with silicon waveguides have shown a hysteretic behavior and optical readout functionality[97]. However, poor ON/OFF signal ratio limits higher-bit densities (i.e. multi-level switching), but a lower power compared to PCMs highlight the potential for femtojoule operation. On the other hand, once the NN is trained a power-optimized bit-density is about 3-bits only. In this context, Photonics Memory, in particular those based on phase change materials and filamentary switching, can help reducing the overall computational cost and overall losses of a multilayer perceptron NN implemented in photonics when performing inference task, by adopting quantization schemes (lower number of bits up to 1), especially in regards of the weights quantization, which could lead towards Photonic Quantized NN (QNN)[98,99]. Additionally, filamentary memristive devices can potentially be employed as electric perceptron mechanism triggered by a plasmonic response[100].

That is, the inference accuracy does only marginally improve for higher bit densities. This is in contrast to online training on the photonic hardware, where higher bit-density reduces the energy cost and training time during gradient descent back-propagation algorithm, while resulting in higher accuracy during inference operations[25]. Nonetheless, Li et al[94], through opportunely shaped pulse, reported a 5 level GST cell, additionally quasi-continuous tuning of GST has been achieved[14] when integrated on a ring resonator.

## 3.2 Neural Network Non-linear Thresholds and Online Training: short-term memory options

Since memristors are challenged by device-to-device variability and require trains of tens of voltage pulses for precise tuning, short-term memory technologies are needed to speed-up both the training and to reduce the energy consumption for on-the-hardware training (offline training, i.e. using GPUs, would be an alternative).

Capacitive-driven devices can be a useful short-term memory solution for integration with photonic platforms, since footprint is not as significant of a concern as in ultra-scaled purely electronic systems. Here, micron-scale capacitors can be fabricated in a stacked configuration atop waveguides, utilizing long-term memristor devices as photonic weights for example (**Fig. 1b**). Unlike capacitive-challenged arrays of electronic memories (either emerging PCMs or classical SRAMs), electro-optic components such as the NL thresholding modulators discussed above are capacitive stand-alone devices, just connected to their respective drivers (e.g. the summation providing photodetector). These modulator-activation functions can respond at <10's ps short delays thus do not slow down the (already fast) optical NN. If longer retention is needed short-term memory elements can be used to bias the modulator's gate temporarily. Such a complex cell could benefit from the advantages of each technology and minimize the drawbacks regarding training.

A novel technology for short-term memory, for instance, is the diffusive memristor[101]; while designed as a selector device compatible with memristive crossbars, the diffusive memristor has the advantage of a metallic filament that can be programmed using low voltages and its memory-state dissolves rapidly (hundreds of μs to ms) afterwards forming nanoparticles. This behavior resembles the $Ca^{2+}$ dynamics in a biological synapse thus providing the short-term memory capabilities required during training. Experiments in plasmonic metasurfaces using silver nano-filaments have shown short-term memory and few volts switching voltage was found similar to that of a diffusive memristors.[102]

The efficient diffusive memristor's dynamic can potentially mimic both short- and long-term plasticity of biological synapses when electrically driven, however when triggered by plasmonic modes this control might not be as straightforward as in the electronic counterpart, also integration of multilevel diffusive memristor in photonics have not be demonstrated, yet. Ultimately, noble metals such as silver are well-known to exhibit plasmonic performances, but their use is restricted in a CMOS foundry as they are contaminants for circuitry, hindering the cointegration process.

## 3.3 Challenges in design of memristor devices for photonic integration

While the integration of non-volatile memristive devices in photonic circuits would be an enabling step towards energy-efficient and NNs capable of nanosecond-short inference tasks, here we address some of the challenges that should be considered in such a roadmap; a typical issue with novel material-chip integration is often a limited reproducibility for larger-scaled PICs.

Unlike NN approaches from the computer science community, which often follow a strategy of adding more neurons to increase inference accuracy and whose NN's approach millions of neurons, recent results on photonic neuromorphic show that; a) 100's of neurons are sufficient to perform smaller inference tasks equally well to electronics[8], and b) the inherent noise of the analog photonic system can be advantageous during training with respect to accuracy. More in details, additional noise in the training phase might be beneficial[103]; a network trained with small amount of noise will be more robust and tolerant to noise during the inference task, and we argue that this could potentially reduce the impact of the quantization error because of discrete weights.

A network trained with small amount of noise will be more robust and tolerant to noise during the inference task, and we argue that could potentially reduce the impact of the quantization error because of discrete weights. Nonetheless, reproducibility and reliability at the device-level is fundamental to ensure performance guarantees at the circuit and system levels during inference, and too high noise will, trivially, impacts accuracy adversely. This problem is related to material engineering, interface control, and optimization of nanofabrication processes. Engineering the material stoichiometry to improve device performance is desirable for a broad range of materials of interest in both memristor and photonics fields. For example, ITO composition can be explored in a holistic fashion using reactive sputtering[104] to carefully control its electrical and optical properties. Similarly, phase-change materials can be tuned to lower the switching energy and improve reliability of amorphization/crystallization. For filamentary memristors, the issue of filament robustness and controllability is driven also by the choice of materials; for example, the electrode material influences the filament shape based on the different free energy of oxide formation or the metal electromigration[105]. Such instabilities lead to noise of the NN. Interestingly, depending on the amount of noise[106], training the NN with the actual system noise of these photonic analog networks results in higher inference accuracy than performing the training 'signal-clean' digitally (i.e. GPUs). Indeed, highlighting material systems, that are jointly shared between memory and photonic technologies could enable a seamless integration pathway (**Fig. 4**).

Here, CMOS compatibility and large-scale integration poses additional material-related constraints, limiting the choice of materials available for electrode, device layer and isolation. For conductive bridge devices, short term transient memory effects on a metasurface can be obtained at voltages as low as 5 mV, yet the thermal noise floor of 26meV at room temperature would dictate an SNR < 1 at the back-end photodetector of the photonic NN[80]. However, long-term memories require often high programming voltages, which reduce the endurance (number of memory's programmable cycles) leading to undesirable electric shorts as the failure mode, which could become extremely critical at high speed. In fact, if the endurance performance of a short-term memory element had a value of $<10^{11}$ cycles and this memory would be used as the charge-storage that controls the NL threshold of the photonic NN, then this memory would fail after just 5 seconds when the inference data input is clocked at 20 GHz. This shows, that some performance parameters do not translate well across these different applications.

However, for usage as NN 'weights' any state-of-the-art memory is already over-performing, given the infrequent updates (which, naturally, depends on the application). Oxide-based memristors have a more tunable filament, at the expense of increased optical insertion loss. Devices based on electro-migration are typically bipolar, requiring both polarities of voltage for programming: one polarity for SET (switching from OFF to ON) and the other polarity for RESET (switching from ON to OFF). The bipolar programming is challenging for all-optical photonic integration, since it is difficult to realize plasmonic optical programming.

A bipolar optical readout would require rectification of the light thus an optical rectenna would be required, yet, so far, optical rectenna technology is still in its infancy exhibiting limited performance[107]. On the other hand, phase-change memories are unipolar, thus all-optical programmability can be achieved, as shown in recent work on all optical STDP plasticity[92]. For long-term memory elements, such neuromorphic technology can interface seamlessly with an optical compute system, thus reducing the overall delay by keeping the signal 'longer' in the optical domain before eventually converting back to electronics via a photodetector. Nevertheless, electro-optic memristors offer advantages due to the integration synergies with circuitry in existing CMOS technology. Therefore, integrated memristor devices exhibiting electrical programmability might be desirable in some integrated photonic systems for offline weight training. These discussion points are qualitatively summarized in Table 2.

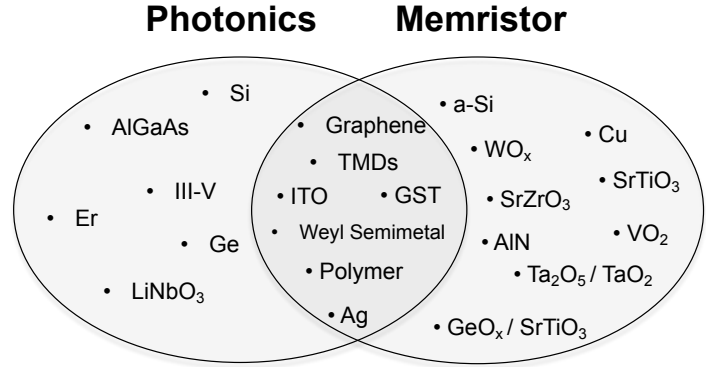

*Figure 4 Representative materials and material systems for photonics and memristors, highlighting the significant overlap. Here we argue that for a multitude of reasons, synergistic opportunities exist when heterogeneously integration memristor materials to act as long and short-term photonic memory elements for neural network weights and nonlinear thresholding, respectively. Such synergies include process integration, reduced derive-to-device capacitance, tailorable bit-density, and programmability (read/write speeds). Prominent options could be ITO and GST as active electro-optic materials for weighting and thresholding, which have both gained semiconductor foundry access recently.*

| Neural Network Function | Mathematical (Perceptron) Function | | HYBRID PHOTONIC-ELECTRONIC NEURON | | | |
|---|---|---|---|---|---|---|
| | | | Photonic Implementation* | Electronic Memory | | |
| | | | | Retention Time | Cycle Endurance | Leakage Current*** |
| Weighting | Dot-product | weighted-addition (multiply-accumulate, MAC) | {Electro-optic Modulator or tunable ring-filters} plus memory**, or, non-volatile material phase-transition | Long-term (dep. on application) | >1-10$^3$ | Important |
| Summation | Summation | | Photodetector or Mx1 fan-in waveguide configuration (coherent system only) | NA | | |
| Nonlinear Activation | Threshold | | Electro-optic Modulator (optional memory), or, all-optical (e.g. photonic bistability, phase change material). *** | Short-term to no memory needed | >10$^{18}$ | Minor Importance |

*Table 2. Qualitative mapping of both the functions of the neural network (NN) and the mathematical model of the perceptron onto photonic-electronic hybrid neurons. The memory is most relevant in providing the charge to store the weights to bias the electro-optic modulators performing weighting. However, the same modulator can also be used as nonlinear activation function. *Assumes analog amplitude signal control (perceptron). Other options such as spiking neurons could also be explored for higher-order neuron signal-shaping using a multitude of partial-differential equation[22]. **if WDM is used. ***Leakage current impacts power consumption of the NN. ***Electro-optic bistability possible to[108].*

# 4 ALL-OPTICAL NONLINEAR ACTIVATION FUNCTION

The above-discussed mechanism for thresholding in a multilayered perceptron relies on photo-detection of the optical power (proportional to the square of the electric field), at the output of the weighted additions for each node to node connection, which is thence send as a gate to an EO modulator. This scheme, however, requires additional laser sources (possibly embedded on-chip), whose intensity is controlled by the electro-optic modulation driven by the photo-generated electrical signal. This represents the main bottleneck of the implementation of the proposed perceptron scheme for achieving NL thresholding, which despite its relatively straightforward implementation and controllability, demands an O-to-E-to-O conversion step at every perceptron; this brings substantial trade-offs in terms of speed and power efficiency of the otherwise intrinsically instantaneous and effortless transmission of the signal through the PIC.

For this reason, an optical nonlinear device with pico-seconds response, with low insertion losses, high modulation range, which doesn't deteriorate, or quench is highly desirable and becomes a prerogative to the implementation of an all optical neuron.
Recently, several steps have been taken in this direction. In reservoir optical computing[109], one of the most accredited optical nonlinearity consists of saturable absorber films, such as graphene layers[110] or based on 2-photon absorption[111]. Other mechanisms are instead based on the nonlinearities of bistable switches[112] and ring resonators[113] have also been investigated. Approaches based on single graphene excitable laser[114] have recently shown significant progress in the field of all optical spiking neural networks. However, the main criticality is still related to the enhancement of the modulation strength, operational speed and the integrability and the reproducibility of such devices in photonic circuit both at the device and system level which represent still an open challenge. Therefore, a combined effort of material engineering and time resolved spectroscopy[17,115–117] aiming to obtain materials with optical nonlinearities which concurrently provides steep modulation range and characterized by fast dynamics could substantially help the advancement of the field of photonics neural networks. According to our predictions[24], a 3-layer fully connected AO-NN would have a delay of about few ps, or $10^{12}$ MAC/s, and up to $10^{17}$ MAC/J efficiency, when performing a standard classification task.

# CONCLUSIONS

In conclusion, the heterogeneous integration of memristive materials and devices in photonic platforms suggest promising performance advantages for photonic-electronic hybrid artificial neural networks, since these two technologies have complementary strengths. The advantage of photonics, for instance, lies in providing a picosecond-short delay between the various network nodes, hence enables highly efficient 'interconnectivity' of such non Von-Neumann compute and information processing architectures. However, photons are challenged to store states, which is where memories come in, such as to provide for the long-term neural network 'weights' (non-volatile) and the nonlinear thresholding or activation function (relatively volatile). In combining the 'best-of-both-words', we see a viable path forward in densely integrating memristive materials particularly with the active photonic components that determine the neural networks governing functions and thus performance.

We find that electro-optic modulators are unique candidates to perform both dot-product 'weighting' as well as 'thresholding', yet vastly different time-scales are required for each, while both share the aim to execute the respective functionality with lowest power consumption. The advantage is strengthened by the fact that the two technologies share a broad range of materials, thus enabling seamless integration and stacking of integrated devices. Indeed, hybrid photonic memristive neuromorphic circuitry could enable systems capable of (sub)nanosecond-fast and energy-efficient inference tasks in trained networks.

On the other hand, promising improvements in the field of non-volatile photonic memory pertains the fabrication of more effective and efficient photonic memories in which a concurrent minimization of the losses and maximization of the modulation keeps the information longer in the optical-domain, i.e. avoids cumbersome O-to-E-to-O conversions. This however, requires wisely engineering the material process, e.g. interfacial PCM (GeTe/$Sb_2Te_3$)[93] and optimized alloy, such as $Ge_2Sb_2Se_4Te_1$[95]. Moreover, the fabrication of frequency selective memories based on PCM can dramatically increase the parallelism and consequently opening new frontiers in optical computing and communication. This vibrant field would also demand for strategies and device configurations which would enable an efficient non-volatile all-optical control in photonic multilevel perceptron, possibly integrated, e.g. photonic crystal cavities[96] or plasmonic slot waveguide[84,118].

Moreover, such platforms could be used to realize new neuromorphic architectures which could rely on hybrid devices based on single photon detection[100] or modulation. Finally, a word on the target applications for photonic neural networks; unlike GPUs which are suitable for big-data and high throughout tasks, photonic neural networks would be quite suitable for those specific tasks that rely on real-time (<microsecond) responses to inference machine learning tasks such as those found in military applications of ranging, synthetic aperture radar, or automated target recognition, to name a few, and a proper task-to-hardware mapping including tradeoffs in network complexity and size against performance (i.e. accuracy, delay) is yet outstanding.